\documentclass[twocolumn,prl,aps,showpacs]{revtex4}

\usepackage{psfrag,graphicx}

\usepackage{dcolumn}
\usepackage{amsmath,amssymb}
\usepackage{bm}
\usepackage[dvips]{color}

\definecolor{mygrey}{gray}{0.35}
\definecolor{mygreen}{rgb}{0.85,1,0.9}
\definecolor{myzard}{cmyk}{0,0,0.05,0}
\definecolor{mywhite}{rgb}{1,1,1}
\definecolor{myred}{rgb}{1,0,0}

\def\R{{\mathbb{R}}}

\def\vec#1{{\bf{#1}}} 

\def\bra#1{\langle#1|} \def\ket#1{|#1\rangle}

\begin{document}

\title[Short Title]{
The Majorization Arrow in Quantum Algorithm Design}

\author{J.I. Latorre$^{\dag}$ and M.A. Mart\'{\i}n-Delgado$^{\ddag}$ 
} 
\affiliation{
$^{\dag}$Dept. d'Estructura i Constituents de la
Mat\`eria, Univ. Barcelona, 08028. Barcelona, Spain.
\\ 
$^{\ddag}$Departamento de F\'{\i}sica Te\'orica I, Universidad Complutense,
28040. Madrid, Spain.  
}

\begin{abstract}
We apply  majorization theory to study the quantum algorithms
known so far and find that there is a majorization principle underlying
the way they operate. Grover's algorithm is a neat instance of this
principle where majorization works step by step  until
the optimal target state is found. Extensions of this situation are
also found in algorithms based in quantum adiabatic evolution and the family 
of quantum phase-estimation algorithms, including Shor's algorithm. 
We state that in quantum algorithms the time arrow is
a majorization arrow. 
\end{abstract}

\maketitle

Majorization is the natural ordering on probability distributions. One probability 
distribution is more uneven than another one when  the
former majorizes  the latter. Furthermore, majorization implies
an entropy decrease, thus
the ordering
concept introduced by majorization is more restrictive and powerful than the one
associated to the Shanon's entropy. The goal of this work is to show
that all known efficient quantum algorithms obey a majorization principle,
in a way to be made precise later.

The classical theory of majorization was first introduced by Muirhead \cite{muirhead}
and later developed by Hardy, Littlewood and P\'olya in their study of symmetric
means \cite{hlp}. Majorization was early studied by economists in the beginning of the
twentieth century in order to formalize the concept of unevenness in the distribution
of income. In 1905, Lorenz pointed out that one distribution can be said to be
more uneven than another precisely when it majorizes the other \cite{marshall-olkin}. 
Likewise, Dalton in 1920 stated his {\em principle of transfers} showing that a 
distribution is less uneven than another if it can be obtained from the other by
transferring some income from a richer to a poorer income-receiver.
Moreover, majorization has found many applications in classical computer science like
stochastic scheduling, optimal Huffman coding, greedy algorithms, etc.

In quantum information theory, majorization 
characterizes when two quantum bipartite pure states can
be connected via Local Operations and Classical Communication \cite{nielsen,vidal}.
This result shows that this connection is indeed possible when there exists
majorization between the vectors of eigenvalues (weights) of the partial 
von Neumann entropies associated to each bipartite state. 
A further application of majorization in quantum information theory
corresponds to the problem of Hamiltonian simulation \cite{ibm}. There,
strong restrictions based on majorization theory limit the possibility to
simulate a proposed quantum evolution from a different given Hamiltonian complemented
with local unitary transformations. Majorization is also present in
quantum measurement theory and in the separability problem.

Majorization is often defined as a binary relation denoted by $\prec$ on vectors
in $\R^d$. We need to fix notations by introducing some basic definitions.

\noindent {\em Definition 1}. For $\vec{x},\vec{y}\in \R^d$,
\begin{equation}
 \vec{x} \prec \vec{y} \ \text{iff} \ 
 \begin{cases}
  \sum_{i=1}^k x_{[i]} \leq \sum_{i=1}^k y_{[i]}, & k=1,\ldots,d-1 \\
  \sum_{i=1}^d x_{[i]} = \sum_{i=1}^d y_{[i]},    &
 \end{cases}
 \label{0}
\end{equation}
where $[z_{[1]}\ldots z_{[d]}]:=\text{sort}_{\downarrow}(\vec{z})$
denotes the descendingly-sorted (non-increasing) ordering of $\vec{z}\in\R^d$.
An immediate consequence is that majorization is a partial order for 
sorted vectors in $\R^d$.

\noindent {\em Definition 2}. If it exists, the least element $x_{\rm l}$ 
(greatest element $x_{\rm g}$) of a partial order like majorization 
is defined by the condition $x_{\rm l} \prec x, \forall x\in \R^d$ 
($x \prec x_{\rm g}, \forall x\in \R^d$).

In this letter we address the following basic problem of elucidating what is the
role, if any,
played  by majorization in the way quantum algorithms operate.
We find, indeed, that there is a majorization principle underlying the way quantum 
algorithms work that we shall now state more precisely.
 Let us denote by $\ket{\Psi_m}$ the pure state
representing the state of the register in a quantum computer at an operating stage 
labeled by $m=0,1,\ldots,M-1$,
where $M$ is the total number of steps of the algorithm.
We can associate 
naturally a set of sorted probabilities $[p_{[x]}], x=0,1,\ldots,2^n-1$
to this quantum state of $n$ qubits in the following way: decompose the register state in the
computational basis 
i.e., $\ket{\Psi_m}:=\sum_{x=0}^{2^n-1} c_x \ket{x}$ with
$\{\ket{x}:=\ket{x_0 x_1 \ldots x_{n-1}}\}_{x=0}^{2^n-1}$ denoting the basis 
states in digital or binary notation, respectively, and $x:=\sum_{j=0}^{n-1} x_j 2^j$.
The sorted vectors to which  majorization theory applies are precisely
$[p_{[x]}]:=[|c_{[x]}|^2]$.
Thus, in quantum algorithms we shall be dealing with probability densities
defined in $\R^d_+$, with $d=2^n$. With these ingredients, our main result
can be stated as follows: in the quantum algorithms known so far,
the set of sorted probabilities $[p^m_{[x]}]$ associated 
to the quantum register at each step $m$  are majorized by the corresponding probabilities of
the next step
\begin{equation}
 [p^m_{[x]}] \prec [p^{m+1}_{[x]}],
 \begin{cases} 
  \forall m=0,1,\ldots, M-2, \\
  x=0,1,\ldots,2^n-1.
 \end{cases}
 \label{00}
\end{equation}
This is a strong result for it means that majorization works locally
in quantum algorithms,i.e.,  step by step,
and not just globally (for the initial and final states).
Our starting point is the majorization analysis of Grover's algorithm \cite{grover1}.

\noindent {\em Grover's algorithm}. 
This quantum algorithm solves efficiently the problem of finding a target item
in a large database.  The algorithm  is based on a
 kernel that acts symmetrically on the subspace orthogonal to
the solution. This is clear from its construction
\begin{equation}
\begin{aligned} 
& K:= U_s U_{y_0}& \\
 U_s:=2\ket{s}\bra{s} -1, &\ \
 U_{y_0}:=1-2\ket{y_0}\bra{y_0} &
\end{aligned}
\end{equation}
where  $\ket{s} :=\frac{1}{\sqrt{N}}\sum_x \ket{x}$ and 
$\ket{y_0}$ is the searched item.

\noindent {\em Theorem:} The set of probabilities to obtain any of the $N$ possible
states in a database is majorized step by step along the evolution of 
Grover's algorithm
when starting from a symmetric state until the maximum probability of success is 
reached.

\noindent {\em Proof.} To prove this result we write    $[p_{[x]}]$  as the set 
of sorted probabilities of
finding the state $\ket{x}$ when performing a measurement. We call
$[p'_{[x]}]$  the set of sorted probabilities after one single application of
Grover's kernel.  The theorem is equivalent
to prove that $[p_{[x]}]\prec [p'_{[x]}]$ until $p_1$, the
probability of finding the correct solution,  reaches its maximum value.

The hypothesis of symmetry imposes that the  probabilities of finding each
of the $N$ outputs at some point during the implementation of
Grover's algorithm can be ordered in the list
\begin{equation}
\textstyle \left[p,{1-p\over N-1},\frac{1-p}{N-1},\dots,\frac{1-p}{N-1}\right],
 \label{3}
\end{equation}
where $p$ is the one associated to the correct output. After one
further action of the kernel these probabilities will be
\begin{equation}
\textstyle 
\left[p',\frac{1-p'}{N-1},\frac{1-p'}{N-1},\dots,\frac{1-p'}{N-1}\right].
 \label{4}
\end{equation}
We first  need to prove that Grover's algorithm increases the probability of success
monotonically, that is $p'>p$, till it reaches
a maximum and then decreases also monotonically. 
This part of the proof relies on the fact that the Grover algorithm 
can be described in a reduced two-dimensional space 
\cite{farhi-gutmann},\cite{family}, which 
follows from the symmetry of the subspace orthogonal to $\ket{y_0}$.
In this case, the dynamics can be reduced to a two-state system, 
$\{ \ket{y_0},
\ket{y_0^\perp} \} $. Grover's kernel on this space acts as a rotation \cite{jozsa}
\begin{equation} 
\textstyle  K=\begin{pmatrix}
  \cos \theta  &-\sin\theta \\
  \sin\theta  &\cos\theta
 \end{pmatrix}
 \label{5}
\end{equation}
where $\cos\theta=1-{2\over N}$. Starting from the  symmetric state
\begin{equation}
\textstyle  \vert s\rangle^{t}=
 \begin{pmatrix}
 \frac{1}{\sqrt N} &
 \sqrt{1-\frac{1}{N}}
 \end{pmatrix},
 \label{6}
\end{equation}
$m$ applications of the kernel lead to
\begin{equation} 
\textstyle  K^m\vert s\rangle= 
 \begin{pmatrix}
  \frac{1}{\sqrt N}\cos m\theta  & -\sqrt{1-{1\over N}}\sin m\theta \\
  \frac{1}{\sqrt N}\sin m\theta  & +\sqrt{1-{1\over N}}\cos m\theta 
 \end{pmatrix}.
 \label{7}
\end{equation}
The projection onto the upper component corresponds to the probability amplitude
which, thus, evolves monotonically until it reaches a maximum.

Returning to the original problem, we can now check that all
probabilities evolve in such a way that majorization works smoothly:
\begin{equation}
\textstyle  \begin{aligned}
  p &\leq  p', \\
  {(N-2)p+1\over N-1} &\leq  {(N-2)p'+1\over N-1}, \\
  \vdots & \\
  {(N-m-1)p+m\over N-1} &\leq {(N-m-1)p'+m\over N-1}.  \\
 \end{aligned}
 \label{8}
\end{equation}
Thus $[p_{[x]}]\prec [p'_{[x]}]$ and (\ref{00}) holds true. $\blacksquare$

Majorization works in a simple way in Grover's algorithm. Nevertheless,
the proof does not hold when the initial distribution of
probabilities is not symmetric in the subspace orthogonal to the
solution. It is indeed easy to find numerical counterexamples to the
majorization principle in absence of symmetry.
We realize that this corresponds to starting with a quantum state $\ket{s}$
whose set of probabilities is the {\em least element} of the majorization
we have introduced to study quantum algorithms. We shall see that this fact
also happens in the rest of algorithms below.

\noindent {\em Quantum adiabatic evolution algorithms}.
Grover's algorithm can be mapped onto the evolution of the homogeneous state
$\vert s\rangle$ into the solution $\vert 0\rangle$ driven by a simple
Hamiltonian \cite{farhi-gutmann}. 
Farhi {\sl et al.} have proposed to use the adiabatic evolution to guarantee
that the system remains in the fundamental state and reaches the
target solution in the end \cite{qadiabatic}. More precisely,  the idea consists
of setting up a Hamiltonian of the form
\begin{equation}
\textstyle  H\left( {t\over T}\right) = \left(1- {t\over T}\right) H_0 +
 {t\over T} H_1
\end{equation}
such that $\vert s\rangle$ is the ground state of $H_0$ and
$\vert 0\rangle$ is the ground state of $H_1$. For large enough $T$, the evolution
will be adiabatic and the system will remain in the ground state all along the
flow. The adiabatic theorem dictates that $T$ must scale as the inverse squared of the
minimum gap of the system. The question we address here
is whether this evolution respects majorization.

\begin{figure}[ht]
 \includegraphics[width=6 cm]{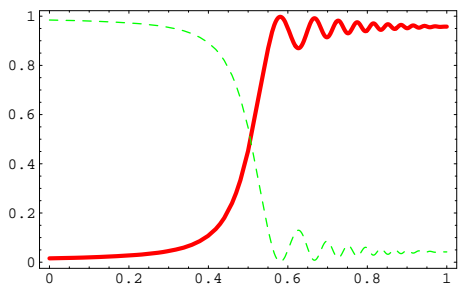}
 \caption{Evolution of the probability of finding the target state
 (bold) and other states (dashed) for $n=6$.}
 \label{grover1}
\end{figure}

Although the system contains $n$ qubits, $2^n$ possible states,
 the adiabatic evolution can be computed
using a subspace if sufficient symmetry is present. 
The simplest example is to consider the  Hamiltonian
\begin{equation}
\textstyle  H\left( {t\over T}\right) = -\vert s\rangle \langle s\vert
  \left(1- {t\over T}\right) -\vert 0\rangle \langle 0\vert {t\over T}
\end{equation}
and the initial state $\vert s\rangle$.
In this particular case, the evolution can be computed using a
reduced two state Hilbert space. More precisely
\begin{equation}
 \vert s\rangle = {1\over \sqrt 2^n} \left( \vert 0\rangle +
    \sqrt{2^n -1}\vert 0^\perp\rangle \right)
\end{equation}
Then the Hamiltonian written in the basis $\{\vert 0\rangle, \vert 0^\perp\}$ reads
\begin{equation}
\textstyle H\left({t\over T}\right)=
 -\left(1-{t\over T}\right)
 \begin{pmatrix}
      1\over 2^n& {\sqrt{2^n-1}\over 2^n}\\
     {\sqrt{2^n-1}\over 2^n}&{2^n-1\over 2^n}
 \end{pmatrix}
  - {t\over T} \begin{pmatrix}
      1 & 0\\
      0& 0
 \end{pmatrix}
\end{equation}
It is possible 
to verify numerically  that when $T\sim 4\  2^n$ the probability follows the
graphic shown in Fig.~\ref{grover1}.
 An argument similar to the previous theorem indicates
that symmetry imposes majorization for the complete set of probabilities. Shorter $T$ lead
to evolutions that do not hit the solution with probability one, while
a larger $T$ smooths this evolution. Once the maximum
is attained, the probabilities oscillate and majorization is obviously lost.

It is worth mentioning that a combination of $H_0$ and $H_1$ chosen as
above but mixed  with no time dependence leads to a Hamiltonian that rotates
the ground state in the manner of the previous theorem. 
Then, the solution is obtained
 in $T={\pi\over 2} 2^{n\over 2}$ with probability 1. This is
precisely the scaling law found in Grover's algorithm.
\begin{figure}[ht]
 \includegraphics[width=4 cm]{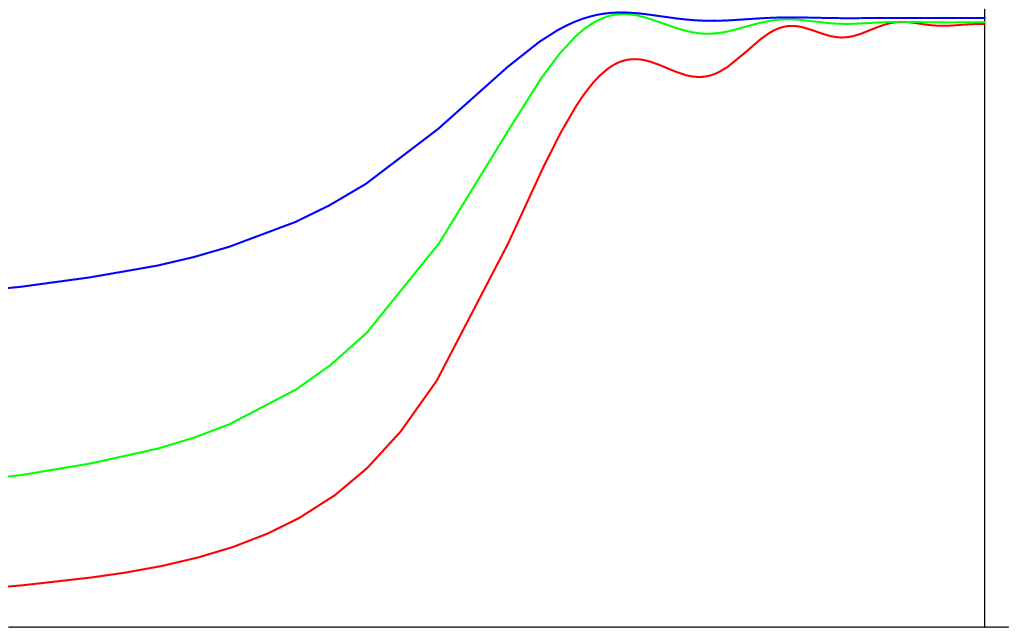}
 \includegraphics[width=4 cm]{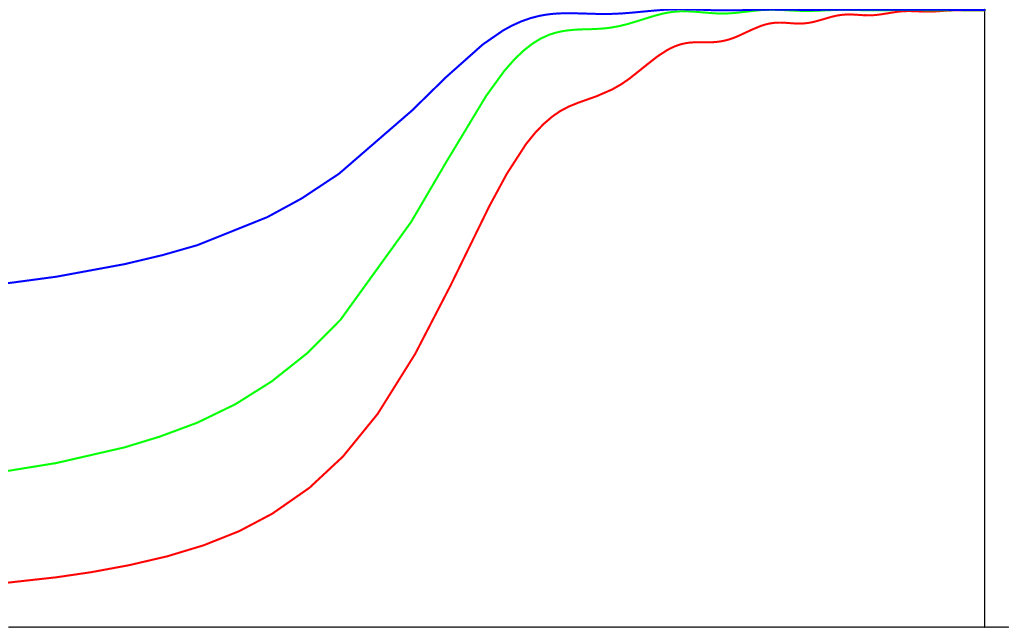} 
 \caption{ Curves for $p_1$, $p_1+p_2$ and $p_1+p_2+p_3$ for $n=4$.
 The Failure of majorization (monotonicity) for fast evolution, $T=4\ 
 2^n$, in the upper curves goes away for 
 slower evolution, $T=7\ 2^n$. }
 \label{grover2}
\end{figure}

A more refined test for the majorization principle
 corresponds to the Hamiltonian evolution proposed by Farhi {\sl et al.} as a
natural starting point for any adiabatic evolution \cite{qadiabatic}.   
Let us consider the following choice
\begin{equation}
 H_0=\sum_{i=1,n} (1-\sigma_x)^{(i)}.
\end{equation}
This Hamiltonian acts as an eraser of information and has the state 
$\vert s\rangle$ as
its ground state. Furthermore, it allows for a decomposition 
of the Hilbert space
into $n+1$ symmetric subspaces. Finding the target instance 
$\vert 0\rangle$ amounts to
solving the dynamical evolution in this  
$(n+1)$-dimensional Hilbert space. Let us denote as
$\vert k\rangle$ as the symmetric space 
with $k$ qubits in the state $\vert 1\rangle$
and the rest in $\vert 0\rangle$. The Hamiltonian becomes
\begin{equation}
 H_0= {n\over 2} I - N,
\end{equation}
where the elements of the symmetric matrix $N$ are given by
\begin{equation}
 \langle i\vert N(i,j)\vert j\rangle=\sqrt{j}\sqrt{n-(j-1)}
 \delta_{i+1,j}.
\end{equation}
A numerical solution of the evolution is now easy to perform. For $T>7
\ 2^n$,
the system indeed evolves along the ground state and majorization holds for the
set of $n+1$ probabilities, as shown in Fig.~\ref{grover2}.
Shorter evolutions perform poorly
and fail to verify the majorization principle. We conclude that
quantum algorithms based on adiabatic evolution naturally fulfill a
majorization
principle provided that the Hamiltonians and initial state are chosen with
sufficient symmetry and the evolution is slow enough.

\begin{figure}[ht]
\psfrag{A}[Bc][Bc][1][0]{a)}
\psfrag{0}[Bc][Bc][0.7][0]{$\ket{0}$}
\psfrag{v}[Bc][Bc][0.75][0]{$\ket{v}$}
\psfrag{h}[Bc][Bc][0.50][0]{$U_H$}
\psfrag{f}[Bc][Bc][0.80][0]{$U_F$}
\psfrag{a}[Bc][Bc][0.70][0]{$U^{2^0}$}
\psfrag{b}[Bc][Bc][0.70][0]{$U^{2^{1}}$}
\psfrag{c}[Bc][Bc][0.70][0]{$U^{2^{n-1}}$}
\psfrag{i}[Bc][Bc][0.70][0]{$\ket{0}+e^{-2\pi i \phi2^{n-1}}\ket{1}$}
\psfrag{j}[Bc][Bc][0.70][0]{$\ket{0}+e^{-2\pi i \phi2^{1}}\ket{1}$}
\psfrag{k}[Bc][Bc][0.70][0]{$\ket{0}+e^{-2\pi i \phi2^0}\ket{1}$}
\psfrag{1}[Bc][Bc][0.75][0]{$t_0$}
\psfrag{2}[Bc][Bc][0.75][0]{$t_1$}
\psfrag{3}[Bc][Bc][0.75][0]{$t_2$}
\psfrag{4}[Bc][Bc][0.75][0]{$t_n$}
\psfrag{5}[Bc][Bc][0.75][0]{$t_{n+1}$}
\includegraphics[width=8 cm]{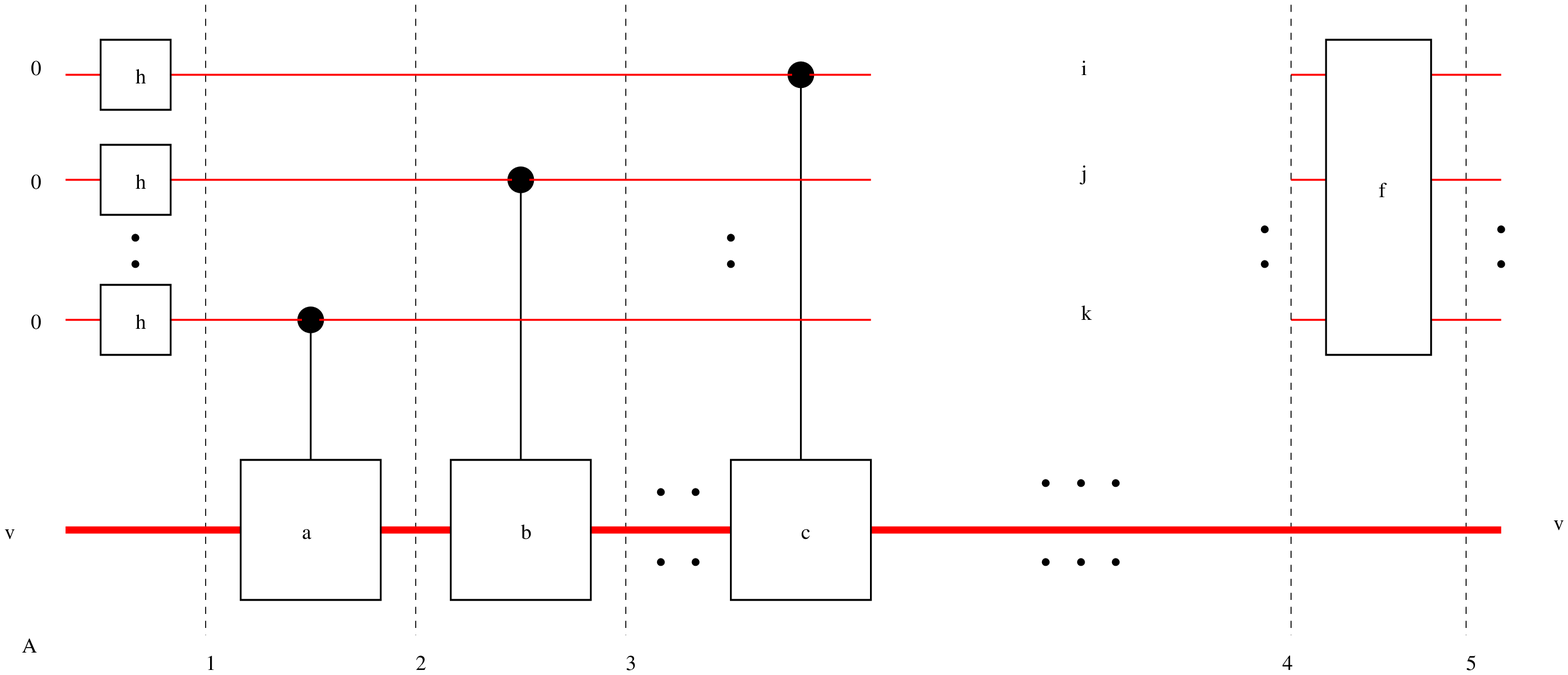}
\psfrag{B}[Bc][Bc][1][0]{b)}
\psfrag{u}[Bc][Bc][0.75][0]{$\ket{v}$}
\psfrag{x}[Bc][Bc][0.65][0]{$U_2$}
\psfrag{y}[Bc][Bc][0.65][0]{$U_3$}
\psfrag{z}[Bc][Bc][0.65][0]{$U_2$}
\psfrag{6}[Bc][Bc][0.75][0]{$t_3$}
\psfrag{7}[Bc][Bc][0.75][0]{$t_{3|1}$}
\psfrag{8}[Bc][Bc][0.75][0]{$t_{3|2}$}
\psfrag{9}[Bc][Bc][0.75][0]{$t_{3|3}$}
\psfrag{p}[Bc][Bc][0.75][0]{$t_{3|4}$}
\psfrag{q}[Bc][Bc][0.75][0]{$t_{3|5}$}
\psfrag{r}[Bc][Bc][0.75][0]{$t_{4}$}
\psfrag{l}[Bc][Bc][0.70][0]{$\ket{0}+e^{-2\pi i \phi2^{2}}\ket{1}$}
\psfrag{m}[Bc][Bc][0.70][0]{$\ket{0}+e^{-2\pi i \phi2^{1}}\ket{1}$}
\psfrag{n}[Bc][Bc][0.70][0]{$\ket{0}+e^{-2\pi i \phi2^0}\ket{1}$}
\includegraphics[width=6 cm]{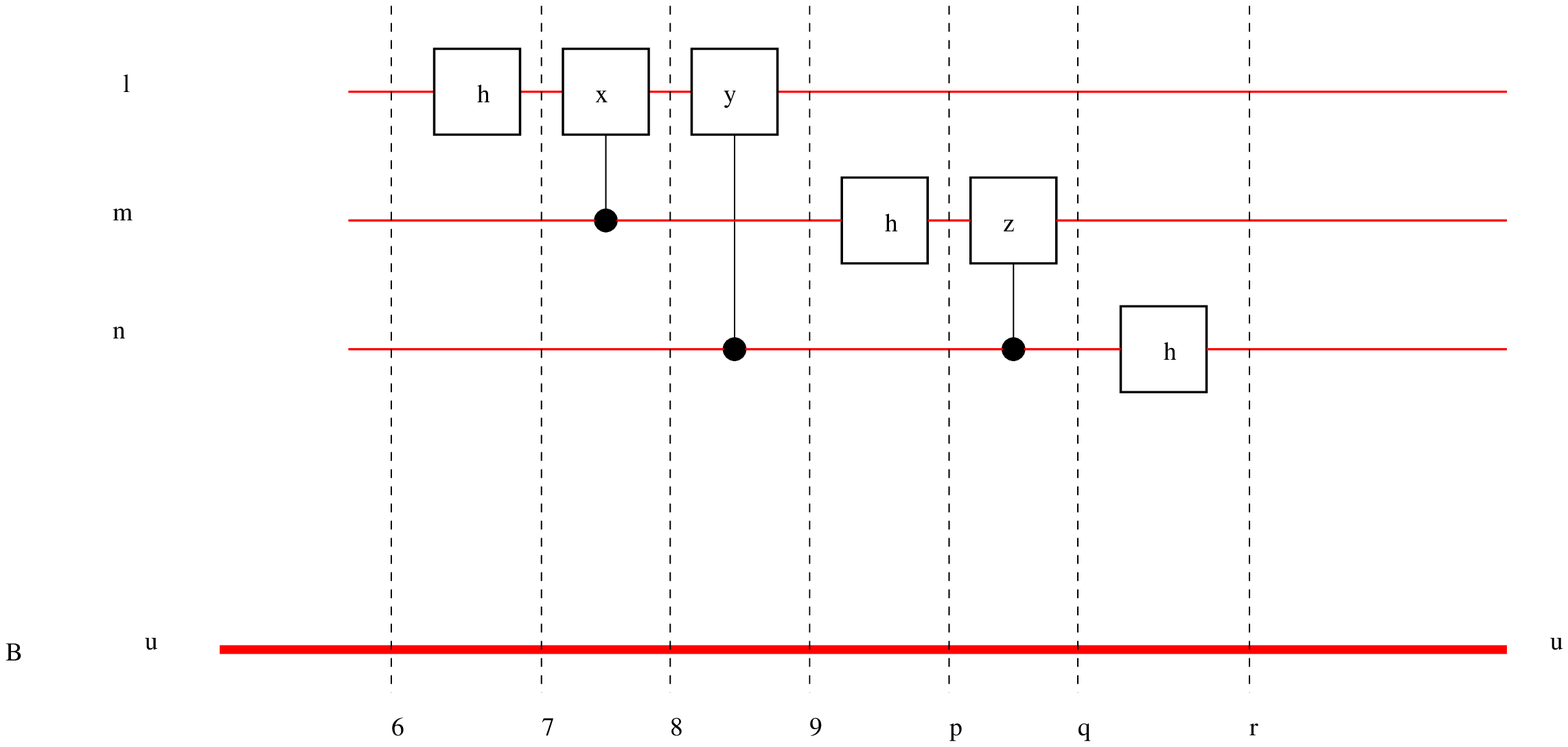}
\caption{a) Quantum circuit implementing the phase-estimation algorithm
constructed from Hadamard gates $U_H$, controlled-$U$ gates
acting as $\ket{0}\bra{0}\otimes 1+\ket{1}\bra{1}\otimes U$,
and the QFT. 
Dashed lines represent time steps for majorization testing.
b) An example of QFT decomposition into elementary
gates for $n=3$ qubits.}
\label{qfase}
\end{figure}

\noindent {\em Quantum phase-estimation algorithms}. These represent
a large family of quantum algorithms that include as particular instances
the order-finding problem, Shor's algorithm \cite{shor}, discrete logarithms, etc.
\cite{revisited}. The basic problem is: given an arbitrary unitary operator
$U$ and one eigenvector $\ket{v}$, estimate the phase $\phi$ of the 
corresponding eigenvalue 
$U\ket{v}:=e^{-2\pi i \phi}\ket{v}, \phi\in[0,1)$, with 
$n$ bits of accuracy. The efficient quantum solution of this problem
can be encoded in the quantum circuit shown in Fig.~\ref{qfase}, and
we shall always refer to this circuit 
when performing the majorization analysis stepwise. 
The algorithm has clearly two parts: i) application of
Hadamard gates $U_H$ and controlled-$U^j$  gates, $j=0,1,\ldots,n-1$;
ii) application of the quantum Fourier transform (QFT) $U_F$.

Part i). The whole quantum register is made up of  first and second
registers. The initialization stage is such that the quantum computer
is in the state $\ket{\Psi_{\rm in}}:=\ket{00\ldots 0} \ket{v}$, where
the first register has been prepared at the state $\ket{0}$ for short,
and the second holds the eigenvector of $U$.
In what follows, we denote by $[p^m_{[x]}]$ the sorted probabilities
distributions of the first register, 
at time steps $m=0,1\ldots,n+1$ that we show in Fig.~\ref{qfase}
as time slices.

Clearly, the probability distribution of $\ket{\Psi_{\rm in}}$ is a
greatest element of the majorization. However, an application of the
Hadamard gates
yields a lowest element as in Grover's algorithm. Thus, our starting
point for majorization is
$\ket{\Psi_0}:=(U_H^{\otimes n}\otimes 1)\ket{\Psi_{\rm in}}=
2^{-n/2} \sum_{x=0}^{2^n-1}\ket{x}\ket{v}.$
Then,
$[p^0_{[x]}]=[2^{-n}],\forall x$.

Next, a series of controlled-$U^{2^j}$ gates
encompassing time steps from $t_1$ to $t_n$ (Fig.~\ref{qfase})
are applied. The outcome of these steps is the factorized state
\begin{equation}
\begin{split}
\ket{\Psi_n}=& 2^{-n/2}
[\ket{0}+e^{-2\pi i2^{n-1}\phi}\ket{1}]\cdots
[\ket{0}+e^{-2\pi i2^{0}\phi}\ket{1}]\\
=& 2^{-n/2} \sum_{x=0}^{2^n-1}e^{-2\pi i x\phi}\ket{x}\ket{v}.
\end{split}
\label{qf2}
\end{equation} 
As the action of these gates
only introduces phases locally in the computational states,
then we obtain again the uniform distributions
$[p^m_{[x]}]=[2^{-n}],\forall x, m=0,1,\ldots,n$.

\begin{figure}[ht]
\psfrag{x}[Bc][Bc][0.75][0]{$x$}
\psfrag{y}[Bc][Bc][0.75][0]{$\sum_{x'=0}^x p^m_{[x']}$}
\psfrag{a}[Bc][Bc][0.65][0]{${\color{red}\bigcirc}\ t_0<t_1<t_2<t_3$}
\psfrag{b}[Bc][Bc][0.65][0]{${\color{green}\square} \ t_{3|1}<t_{3|2}<t_{3|3}$}
\psfrag{c}[Bc][Bc][0.65][0]{${\color{blue}\diamondsuit} \ t_{3|4}<t_{3|5}$}
\psfrag{d}[Bc][Bc][0.65][0]{$\triangle \ t_4$}
 \includegraphics[width=6 cm]{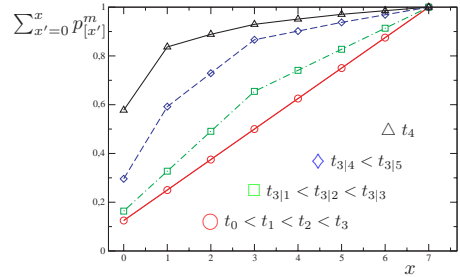}
\vspace{-0.8 cm}
\caption{Lorenz diagram (partial probability sums) for the quantum
phase-estimation algorithm with $\phi=0.2$ and $n=3$ qubits
as in Fig.~\ref{qfase}. It shows how majorization works along
the time arrow
${\color{red}\bigcirc}\rightarrow {\color{green}\square} \rightarrow
{\color{blue}\diamondsuit} \rightarrow \triangle$.}
 \label{lorenz}
\end{figure}

Part ii). Although the local phases in $\ket{\Psi_n}$
do not play any role in majorization, so far, they become relevant
when combined with the application of the QFT on the
first register,
due to interference of quantum amplitudes.
The state after time step $t_{n+1}$ (Fig.~\ref{qfase}) is
\begin{equation}
\ket{\Psi_{n+1}}:= (U_F\otimes 1)\ket{\Psi_{n}}
= 2^{-n} \sum_{x,y=0}^{2^n-1}e^{-2\pi i x(\phi-y/2^n)}\ket{y}\ket{v}.
\label{qf3}
\end{equation}
Now,
$p^{n+1}_{[y]}:=|2^{-n} \sum_{x=0}^{2^n-1}e^{-2\pi i x(\phi-y/2^n)}|^2$
majorizes the least element distribution at step $m=n$.
Interestingly enough, there is a stronger majorization working stepwise
when the QFT is applied by means of its canonical decomposition in terms
of $n$ Hadamard and $n(n-1)/2$ controlled-phase gates \cite{coppersmith}.
For concreteness, we show such decomposition in Fig.~\ref{qfase}b) for
$n=3$ qubits and with the corresponding time slices (majorization checkpoints).
The proof of this result relies on the recursive application
of the following inequalities
\begin{equation}
\left|\frac{1}{\sqrt{2}} (1\pm e^{2\pi i\alpha_{\pm}(y,\phi)})\right|^2\geq 1,
\begin{cases}
\alpha_+\in [0,\frac{1}{4}],[\frac{3}{4},1], \\
\alpha_-\in [\frac{1}{4},\frac{3}{4}],
\end{cases}
\label{qf4}
\end{equation}
where, at each step, $\alpha_{\pm}$ depends on $y,\phi$ in a computable way
\cite{note}.
To illustrate this fact, we show in Fig.~\ref{lorenz} a numerical plot for
$n=3$ qubits in the form of a Lorenz diagram: partial probability sums
vs. $x$, for each time step.
\vspace{-0.8 cm}
Therefore, as a consequence of our analysis we find that the majorization
principle is working locally in algorithms like order-finding
$a^r=1 \ {\rm mod}\ N$, where
the unitary operator is given by $U\ket{x}:=\ket{a x \ {\rm mod}\ N}$
and $\phi=1/r$; Shor's algorithm, where order-finding is used combined
with controlled-$U$ gates implementing the modular exponentiation;
Chuang's algorithm for quantum clock synchronization, where
$U:=U_{\rm cnot}U_{\rm TQP}U_{\rm cnot}$ and $U_{\rm TQP}$
is the so called
Ticking Qubit Protocol \cite{qclock}; etc.

{\em Conclusion}.
Efficient quantum algorithms are
scarce as compared with their classical counterparts, suggesting that
we are missing the basic principles for quantum algorithm design \cite{rmp}.
In this note,  we have produced evidence for the general idea
 that there is a majorization principle acting step by step during
the time evolution in efficient quantum algorithms.
We may say that majorization is a sort of driving force for
such algorithms.
Learning to tame majorization may
be useful for devising quantum algorithm design.
When majorization is not at work, the quantum algorithm is neither efficient
nor successful.

\noindent {\em Acknowledgments}. We are grateful to G. Vidal for
introducing us to majorization theory ideas and for his advice.  
We acknowledge financial support
from the projects:  AEN99-0766, 1999SGR-00097, IST-1999-11053 and PB98-0685.


\begin{references}
\bibitem{note} For $n=2$ qubits, the proof is simple but lengthy. 
Then, the extension
to an arbitrary number of qubits follows.
\end{references}
\end{document}